\documentclass[aps,prl,showpacs,floats,twocolumn,floats,superscriptaddress,floatfix]{revtex4-1}
\usepackage{graphicx}
\usepackage{bm}
\usepackage{amsfonts}
\usepackage{multirow}
\usepackage{color}
\usepackage{amsmath}    % need for subequations
\usepackage{epsfig}
\usepackage{subfigure}  % use for side-by-side figures
\usepackage{hyperref}   % use for hypertext links, including those to external documents and URLs
\usepackage{bm}
\usepackage{amssymb}
\newcommand{\kg}{{K_{\scriptscriptstyle \mathrm{G}}}}

\newcommand{\pkg}{\mathrm{P}_{\!\!\! _{K_{\scriptscriptstyle
        \mathrm{G}}}}}

\newcommand{\ue}{\mathrm{e}}
\newcommand{\ui}{\mathrm{i}}

\newcommand{\ictsaddress}{International Centre for
  Theoretical Sciences, Tata Institute of Fundamental Research,
  Bangalore 560089, India}
\begin{document}
\title{The Onset of Thermalisation in Finite-Dimensional Equations of Hydrodynamics: Insights from the Burgers Equation}
\author{Divya Venkataraman\email{ds}}
\email{Presently at the Department of Mathematics, Institute of
Chemical Technology, Mumbai, India}
\email{vdivya8@yahoo.co.in}
\address{\ictsaddress}
\author{Samriddhi Sankar Ray}
\email{samriddhisankarray@gmail.com}
\address{\ictsaddress}
%\corres{Samriddhi Sankar Ray\\
%\email{samriddhisankarray@gmail.com}
\begin{abstract}

Solutions to finite-dimensional (all spatial Fourier modes set to zero beyond a finite 
wavenumber $K_G$), inviscid equations of hydrodynamics at long times are known to be at 
variance with those obtained for the original infinite dimensional partial differential 
equations or their viscous counterparts. Surprisingly, the solutions to such Galerkin-truncated 
equations develop sharp localised structures, called {\it tygers} [Ray, et al., Phys. Rev. E {\bf 84}, 
016301 (2011)], which eventually lead to completely thermalised states associated with an 
equipartition energy spectrum. We now obtain, by using the analytically tractable Burgers equation, 
precise estimates, theoretically and via direct 
numerical simulations, of the time $\tau_c$ at which thermalisation is triggered and show that 
$\tau_c \sim \kg^\xi$, with $\xi = -4/9$. Our results have several implications including for the 
analyticity strip method to numerically obtain evidence for or against blow-ups of the 
three-dimensional incompressible Euler equations.  

\end{abstract}
%\subject{fluid mechanics, applied mathematics, chaos theory}
\maketitle
\section{Introduction}

A microscopic understanding of turbulent flows have been amongst the most
challenging problems in statistical physics for many years. Central to this
challenge is adapting well-developed tools of statistical mechanics for
dissipative and out-of-equilibrium turbulent flows. The early efforts in this
direction, due to E. Hopf~\cite{hopf} and T. D. Lee~\cite{lee}, treated the
ideal (inviscid) equations of hydrodynamics as a finite-dimensional,
Galerkin-truncated system and obtained equipartition solutions 
with an energy
spectrum, in three dimensions, $E(k) \sim k^2$, at variance with the Kolmogorov
result $E(k) \sim k^{-5/3}$ for dissipative turbulent flows~\cite{K41}.

A major stumbling block in developing a framework to understand
out-of-equilibrium  turbulent flows in the language of classical equilibrium
statistical mechanics is that a microscopic, Hamiltonian formulation of fluid
motion with statistically steady states characterised by an invariant Gibbs
measure inevitably fails.  This is because a self-consistent macroscopic point
of view will invariably lead to an irreversible energy loss and thus a
dissipative hydrodynamic formulation. However, since the pioneering work of
Hopf and Cole, and despite the successes in adapting statistical mechanics
methods in two-dimensional turbulence~\cite{kraic67}, the precise relation
between statistical physics and turbulence remains an open question.

An important breakthrough in understanding this surprising connection came in
the work of Majda and Timofeyev~\cite{majda00} on the one-dimensional Galerkin
truncated Burgers equation which, while exhibiting intrinsic chaos, has
nevertheless a compact equilibrium statistical physics description. The
solution to this equation was shown to thermalise, with energy equipartition,
in a finite time. Subsequently, Cichowlas {\it et al.}, in
Ref.~\cite{brachet05} discovered through state-of-the-art direct numerical
simulations (DNSs) of the finite-dimensional, truncated, incompressible Euler
equations that the solutions in a finite time start thermalising. This process
of thermalisation begins at the largest wavenumber of the system -- the
truncation wavenumber $\kg$ (such that all modes with wavenumbers greater than 
$\kg$ are set to zero) -- and with time starts extending to smaller and
smaller wavenumbers until eventually one obtains an energy spectrum $E(k) = k^2$
for all wavenumbers $1 \le k \le \kg$. Curiously, it was observed that at
intermediate times the energy spectrum showed a transient Kolmogorov scaling
$E(k) \sim k^{-5/3}$ at smaller non-thermalised wavenumbers and a scaling of
$E(k) \sim k^2$ for higher wavenumbers all the way upto $\kg$. This seminal work
thus provided the first numerical evidence of the co-existence of equilibrium
micro-states along side a Kolmogorov-like turbulent cascade in inviscid,
finite-dimensional equations of hydrodynamics (see also Ref.~\cite{kraich89}).  

A second, recent, breakthrough in understanding the interplay between
equilibrium statistical mechanics and turbulent flows came through the
development of the method of fractal Fourier decimation introduced in
Ref.~\cite{frisch12}. This novel method, which allows microsurgeries in the
triadic interactions of the non-linear term, was used to show, theoretically
and numerically, that there exists special dimensions where fluxless,
equilibrium solutions coincide with the Kolmogorov spectrum~\cite{frisch12}
(see also Ref.~\cite{lvov02}).  Subsequently  this method was used in several
other studies~\cite{buzzicotti1,biferale1,biferale2,lagnjp} to
understand triadic interactions {\it inter alia} intermittency, equilibrium
solutions and turbulence. 

As a result of these very recent
developments~\cite{brachet05,ray11,frisch12,majda00,lvov02,NazarenkoPRB,Colm,Feng}
(see also Ref.~\cite{ray-review} for a recent review), the last few years have
seen a furthering in our understanding of the possible existence of equilibrium
solutions in inviscid equations of hydrodynamics.  Specifically, equipartition
solutions to the Galerkin-truncated Gross-Pitaevskii~\cite{GP},
magnetohydrodynamic~\cite{GKMHD11}, Burgers ~\cite{majda00,ray11}, and Euler
equations~\cite{brachet05,GK09} have been studied extensively in recent years.
Alongside the very important theoretical underpinnings of such studies,
thermalised or partially thermalised states have been
shown~\cite{frisch08,frisch13,banerjee14} to be a possible explanation of the
ubiquitous bottleneck~\cite{bottleneck} in the energy spectrum of turbulent
flows.

Despite the rapid advances in this field~\cite{ray-review}, an important
question remains unanswered. It has been shown in earlier
studies~\cite{brachet05,ray11,ray-review} that inviscid, truncated systems
thermalise in finite times.  The reason why one obtains thermalised states is
of course well known~\cite{majda00,ray11}: Galerkin-truncated (which we define
precisely later) equations of ideal hydrodynamics (such as the Euler or the
inviscid Burgers equation) are conserved dynamical systems with Gibbsian
statistical mechanics~\cite{brachet05,ray11,kraich89}.  A few years ago the
first explanation of how such systems thermalise, through resonant
wave-particle interactions mediated by structures called {\it tygers} was given
in Ref.~\cite{ray11}.  However a theoretical or numerical estimate of the
time-scale at which thermalisation sets in has proved elusive.  In this paper
we address this question and show through theoretical arguments,  which are
substantiated by detailed numerical simulations, that the time $T_c$ at
which thermalisation sets in scales as $T_c \sim \kg^{-4/9}$, where $\kg$ is
the Galerkin truncation wavenumber (defined precisely below) in the problem.

There is one additional, applied reason for investigating the issue of the
onset of thermalisation in such systems.  Spectral and pseudospectral methods, 
especially with the advent of fast Fourier transform routines, are extremely 
precise techniques for numerical solutions of the nonlinear partial differential 
equations of hydrodynamics~\cite{dns}. By definition such methods are limited by finite 
resolutions and hence are finite-dimensional. Therefore such numerical methods 
nearly always end up solving the Galerkin-truncated variant of the actual equation 
of hydrodynamics. In viscous calculations, such as the Navier-Stokes or the viscous 
Burgers equations, the difference between the true solution of the equations and 
its Galerkin-truncated variant is often imperceptible. However in numerical 
studies of the blow-up problem~\cite{blow-up} for inviscid equations the onset of thermalised 
states -- which are not admitted in the original infinite dimensional partial differential equation -- 
can have grave consequences on the interpretation of numerical results. 

\section{Galerkin Truncation}

Given the formidable theoretical (and even numerical) difficulties associated with the 
incompressible, three-dimensional Euler equations, it seems that a convenient starting point 
to explore the onset of thermalisation in inviscid, finite-dimensional equations of 
hydrodynamics is the one-dimensional Burgers equation. Given its analytical and numerical
tractability, the Burgers equation has often been used with great success to establish 
or disprove conjectures for problems pertaining to the Euler or Navier-Stokes equations.
(We refer the reader to Refs.~\cite{burgreview1,burgreview2} for a review of the Burgers 
equations and its many applications.) We should point out that even in the relatively simpler 
problem of the inviscid, truncated Burgers equation, the limit $\kg \to \infty$ is not a trivial 
one. Indeed there are examples of energy-conserving perturbations to the inviscid Burgers equation 
which do not necessarily converge (in a weak sense) to the inviscid limit~\cite{no-conv}.

\begin{figure*}
\includegraphics[width=0.24\linewidth]{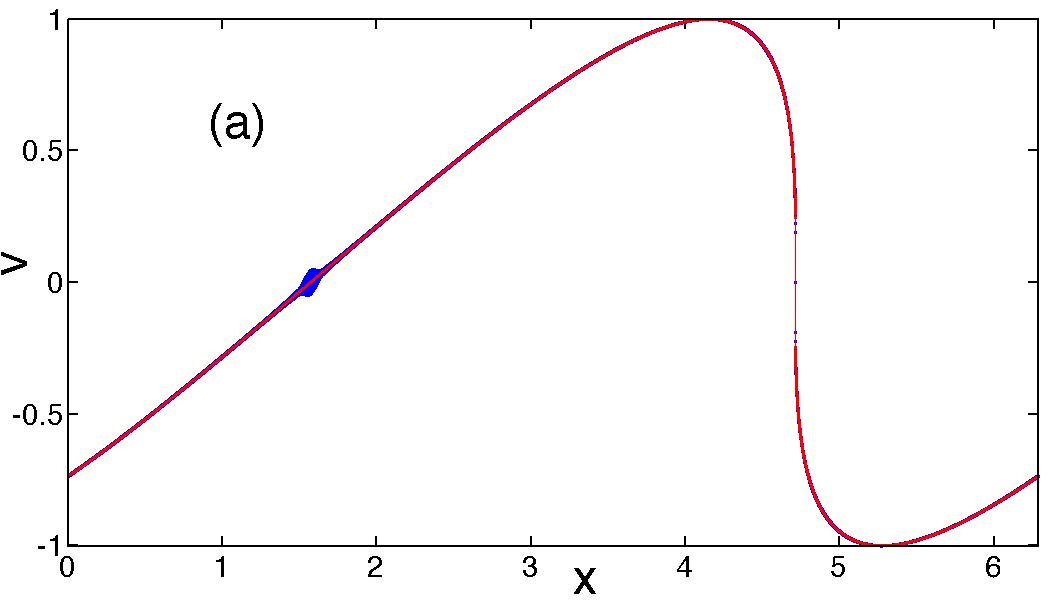}
\includegraphics[width=0.24\linewidth]{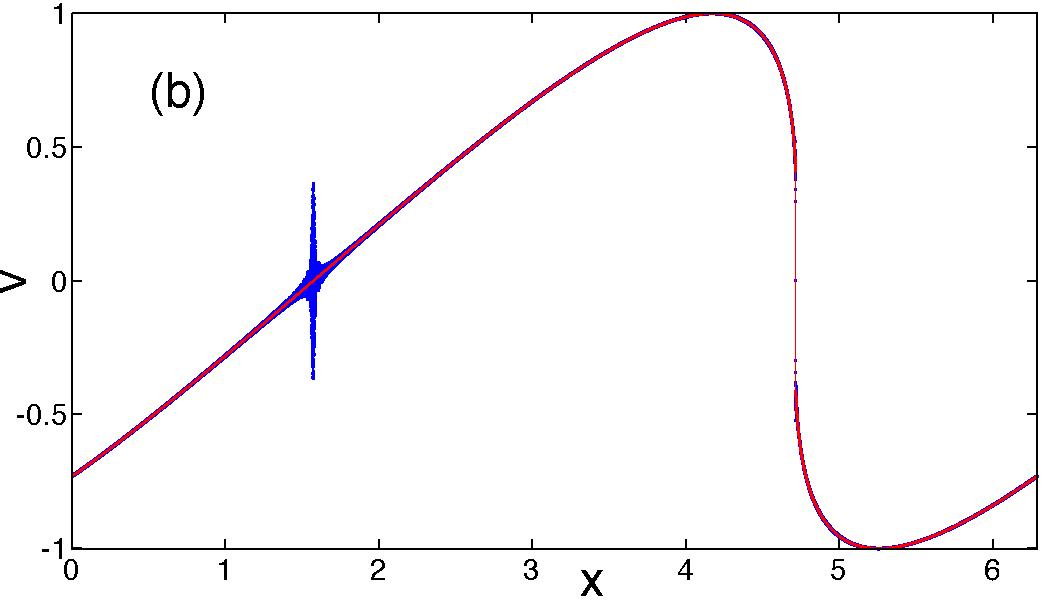}
\includegraphics[width=0.24\linewidth]{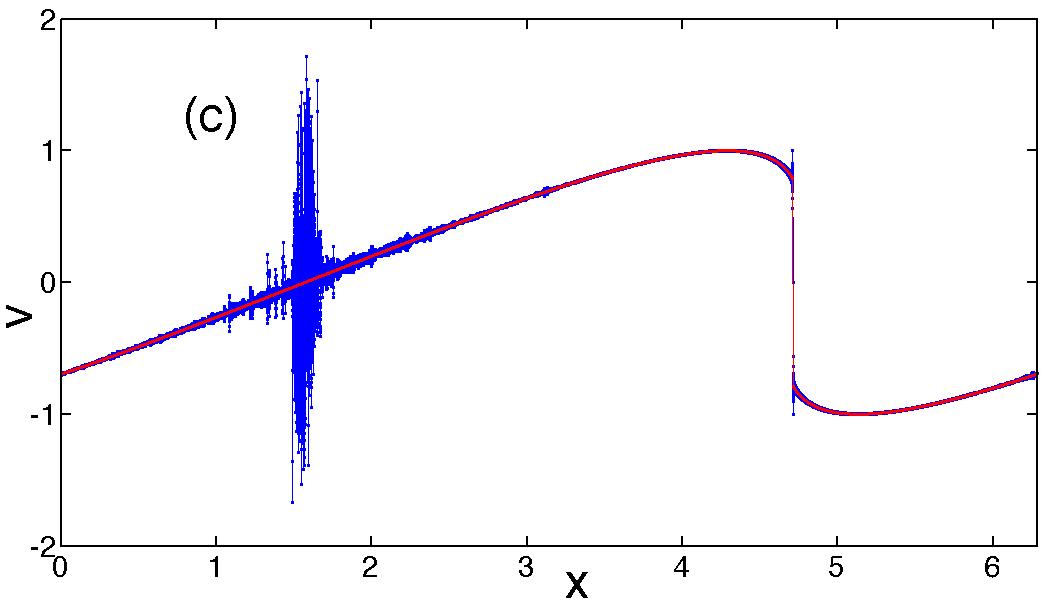}
\includegraphics[width=0.24\linewidth]{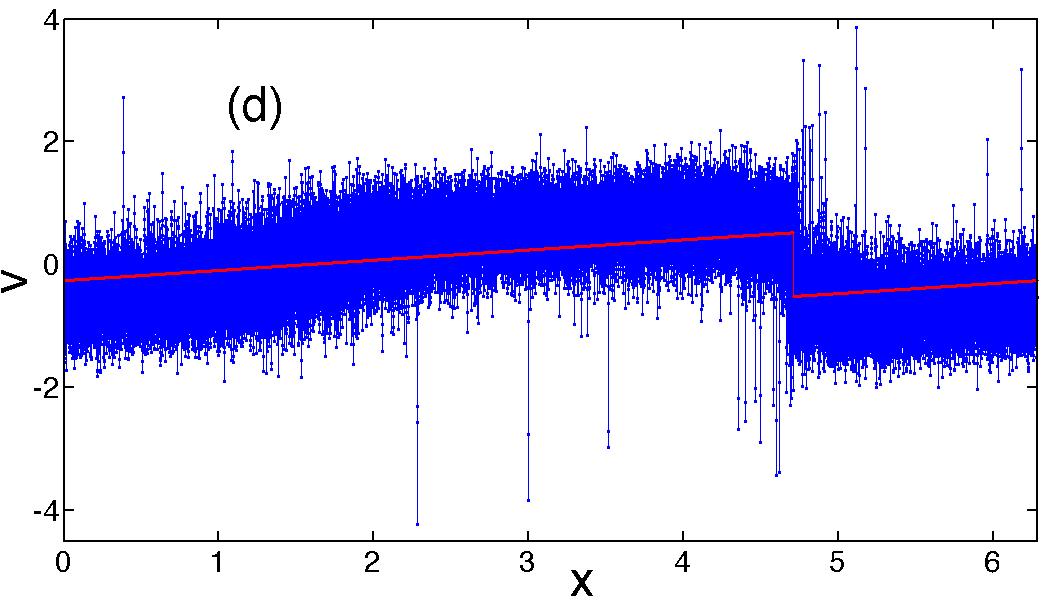}
\caption{Solution of the Galerkin-truncated Burgers equation~\eqref{gtburgers} $v$ (in blue) and the un-truncated Burgers 
equation~\eqref{burgers} (entropy solution ) $u$ 
(in red) at different times for the initial condition $v_0 = u_0 = \sin (x + \phi)$ at different times 
$t \gtrsim t_\star$. At early times (a) $t = 1.01$ the discrepancy between the two is small and localised  around 
$x = x_{\rm tyger}$ 
where the velocity is dentical to that of the shock. This discrepancy, the tyger, grows (still remaining localised in space) in time 
as shown in (b) with $t = 1.03$. At a later time (c) $t=1.15$, the symmetric, localised structure becomes asymmetric and starts 
spreading away from $x_{\rm tyger}$. Even at this time the truncated solution is able to duplicate the 
entropy solution at regions far away from $x_{\rm tyger}$ and in particular is able to capture the location and 
amplitude of the shock. At much later times, (d) $t = 5.0$, the solution $v$ thermalises and becomes Gaussian, 
does not display the shock,  and shows no similarity with the entropy solution $u$. This set of representative plots 
showing the birth and growth of tygers followed by the onset and eventual thermalisation of the solution were obtained 
from simulations with $\kg = 5000$ and $N = 2^{16}$.}
\label{time-evolve}
\end{figure*}

Thus, we begin with the one-dimensional inviscid Burgers equation
\begin{equation}
 \frac{\partial u}{\partial t} + \frac{1}{2}\frac{\partial u^2}{\partial x} = 0
\label{burgers}
\end{equation}
augmented by the initial conditions $u_0(x)$ which are typically a combination of trigonometric 
functions containing a few Fourier modes. Without any loss in generality one can choose $u_0(x) = \sin (x + \phi)$; 
where $\phi$ is some random phase. Since we work in the space of $2\pi$ periodic solutions, we can 
expand the solution of Eq.\eqref{burgers} in a Fourier series
\begin{equation} 
u(x) = \sum_{k= 0, \pm 1, \pm 2 \ldots} \ue ^{\ui kx}\hat u_k .
\label{fourier}
\end{equation}
This allows us to naturally define the Galerkin projector $\pkg$ as a low-pass filter
which sets all modes with Fourier wavenumbers $|k| > \kg$, where $\kg$ is a positive (large) 
integer, to zero via 
\begin{equation} 
\pkg u(x) =   \sum_{|k| \le \kg} \ue ^{\ui kx}\hat u_k.
\label{defpkg}
\end{equation}

These definitions allow us to write the Galerkin-truncated inviscid Burgers equation as  
\begin{equation} 
 \frac{\partial v}{\partial t} + \pkg\frac{1}{2}\frac{\partial v^2}{\partial x} = 0;
\label{gtburgers}
\end{equation}
the initial conditions $v_0 =\pkg u_0$ are similarly projected onto the subspace spanned by $\kg$. This defines 
the Galerkin-truncated velocity $v$ of the Burgers equation. For the three-dimensional Euler equations, the same definition  follows 
{\it mutatis mutandis}.

The inviscid Burgers equation ~\eqref{burgers} conserves all moments of the
velocity field; by contrast the Galerkin-truncated version of it
~\eqref{gtburgers} retains only the first three moments of $v$, and
in particular the energy.  Numerically, however, the dynamics of the
Galerkin-truncated Burgers captures rather well the blowing up (with smooth
initial conditions) of the gradient of the solution to the inviscid Burgers
partial differential equation in a finite time $t_\star$. Indeed the cubic-root
singularity (preshock), at $t_\star = \frac{1}{\rm {max} [\partial_x u]}
\approx \frac{1}{\rm {max} [\partial_x v]}$, in $u$ \eqref{burgers} is also
seen in the solution $v$ of the Galerkin-truncated equation
\cite{ray11,burgreview1,burgreview2,fournierfrisch}. Theoretically, the
solution to ~\eqref{burgers} for times greater than $t_\star$ is obtained by
adding a tiny viscous dissipation term $\nu\frac{\partial^2 u}{\partial x^2}$,
with $\nu \to 0$ (the inviscid limit), which yield, depending on the initial
conditions, finitely many shocks for times $t > t_\star$~\cite{hopf}. 
This generalised solution, which converges weakly to the inviscid Burgers equation~\eqref{burgers}, 
is characterised by the dissipative anomaly: energy dissipation $\epsilon$ remains finite (with an 
associated non-conservation of the total energy) as $\nu \to 0$.  
This is very different to the dynamics of the Galerkin-truncated
equation \eqref{gtburgers} whose solution $v$ stays smooth, conserves energy for all times $t > t_\star$ and 
later thermalises.

\section{Simulation Details}

We perform extensive and state-of-the art simulations to obtain solutions for $u$~\eqref{burgers} and 
$v$~\eqref{gtburgers} for all times. 

The true or entropy solution to the inviscid Burgers equation ~\eqref{burgers} is obtained by the method of Fast Legendre Transform, 
which takes the limit $\nu \to 0$, developed 
by Noullez and Vergassola~\cite{noullez} (see also Refs.~\cite{bec00,mitra05}). This method takes advantage of the 
fact~\cite{burgreview1,burgreview2} that the velocity potential $\psi$ defined via $u = -\frac{\partial \psi}{\partial x}$ 
is constrained by the maximum principle such that 
\begin{equation}
\psi(x,s) = \max_y \left[\psi(y,t) -
\frac{(x-y)^2}{2\, (s -t)} \right]; 
\label{eq:psi}
\end{equation}
where $s > t$. We typically use the number of collocation points $N = 2^{14}$ or $2^{16}$ and choose a time step $\delta t$ large enough 
for a Lagrangian particle\footnote{The inviscid Burgers dynamics can be mapped onto a Lagrangian system of colliding 
particles which form shocks without crossing each other~\cite{burgreview1,burgreview2}.} to move by a distance equal to or larger 
than the grid spacing $\delta x$, and smaller than all other time scales associated with the dynamics. 

We, of course, use a different strategy to solve the Galerkin-truncated equation~\eqref{gtburgers}. 
We use a pseudo-spectral method~\cite{dns} coupled to a 
fourth-order
Runge--Kutta time marching with the total collocation points $N = 2^{14}$ or $2^{16}$ as before. We use 
different values of $\kg$, ranging from 700 to 20,000. Our time step $\delta t$ varied 
from $10^{-4}$ for $\kg \le
5,000$ to $10^{-5}$ for $\kg >5,000$. Full dealiasing was ensured in this problem via $\kg$.

\begin{figure}
\includegraphics[width=1.05\columnwidth]{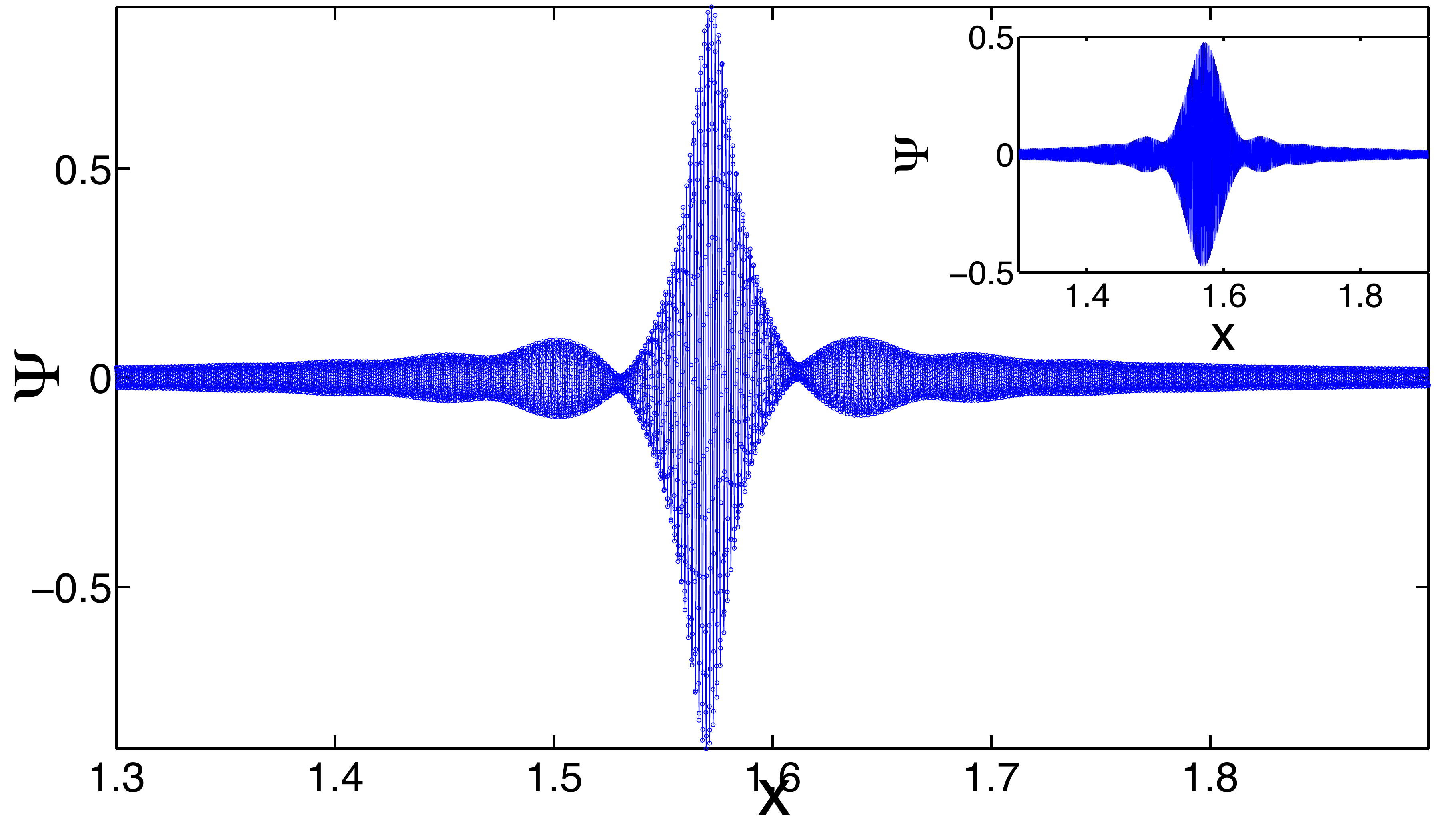}
\caption{ A representative zoomed-in plot of the discrepancy $\Psi = v - u$, around $x_{\rm tyger}$ for $\kg = 4000$ just 
when the bulge becomes asymmetric $t = T_c$ and (inset) when it is still symmetric $t < T_c$. The tyger, apart from being localised, 
is monochromatic with a single wavenumber $\kg$ before $T_c$ (inset)~\cite{ray11}; for $t \ge T_c$, other harmonics smaller than $\kg$ are 
generated as well.}
\label{bulge}
\end{figure}

\section{Results}
\subsection{Tygers and Thermalisation}

We begin by performing direct numerical simulations of
equations~\eqref{burgers} and \eqref{gtburgers}, without any loss of
generality~\cite{ray11,ray-review}, with initial conditions $u_0(x) = v_0(x) =
\sin (x + \phi)$, where $\phi = 0.7$ is a phase which shifts the location of
the cubic singularity away from $x = \pi$. For such an initial condition, it is
easy to show that $t_\star = 1.0$~\cite{burgreview1,burgreview2}.  The time
evolution of the solution to the untruncated Burgers equation $u(x)$ and the
truncated Burgers equation $v(x)$ are shown in Fig.~\ref{time-evolve}. For
times $t \lesssim t_\star = 1.0$, $v = u$ $\forall x$ as shown before
\footnote{Actually, as shown by Ray, {\it et al.}~\cite{ray11}, this
discrepancy is undetectable upto a time $t = t_\star -
\mathcal{O}\left (\kg^{-2/3}\right )$.}.  At $t = t_\star$ (in this case $t_\star = 1.0$)
and thereafter the discrepancy between the entropic solution $u(x)$ (in red)
and $v(x)$ (in blue) is large as seen in Fig.~\ref{time-evolve}. At early
times, $t\gtrsim t_\star$, this discrepancy is localised
(Figs.~\ref{time-evolve}a and \ref{time-evolve}b);  at later times
(Fig.~\ref{time-evolve}c) the solution to the Galerkin truncated equation
starts deviating strongly from the entropy solution till at very large times
it thermalises and shows a white-noise behaviour markedly different from the
saw-tooth entropic solution (Fig.~\ref{time-evolve}d). 

A useful framework to study the departure from the entropy solution to the
thermalised solution is via the discrepancy $\Psi = v - u$. For $t \lesssim
t_\star$, $\Psi = 0$.  However, as shown in Ref.~\cite{ray11}, at $t =
t_\star$, at points which have the same velocity as the shock(s), $\Psi \neq 0$
and a symmetric, localised, monochromatic bulge, called {\it tyger} by Ray,
{\it et al.}~\cite{ray11}, forms as shown in Fig.~\ref{bulge} (inset). As time
evolves, this bulge grows in amplitude $\alpha$ and width $\beta$, becomes
asymmetric (Fig.~\ref{bulge}, at $t = 1.07$), collapses, delocalises, and eventually  invades
the whole $2\pi$-domain (Figs.~\ref{time-evolve}c and \ref{time-evolve}d).

We now know that  finite-dimensional equations of hydrodynamics thermalise 
through wave-particle resonances. Such resonances at early times
manifest themselves as localised structures at the instant $t_{\rm G} \lesssim
t_\star$, when complex singularities are within one Galerkin wavelength $\lambda_{\rm
G} = \frac{2\pi}{\kg}$ of the real domain. This was shown to be true for both
the compressible Burgers equation as well as the incompressible Euler
equations~\cite{ray11,ray-review}.  Indeed the scaling properties of the early
tygers at $t = t_\star$ were derived in Ref.~\cite{ray11} and in particular the
amplitudes and widths were shown to scale as $\alpha \sim \kg^{-2/3}$ and
$\beta \sim \kg^{-1/3}$, respectively. Finally it was shown through detailed
numerical simulations that in a short time these monochromatic, with wavelength
$\lambda_{\rm G}$, localised, symmetric tygers become skewed, {\it collapse},
and spread throughout $0 \le x \le 2\pi$, generate other harmonics, and
eventually result in a thermalised solution with energy equipartition $E(k)
\sim k^0$. However the critical question of the time $T_c$ when
thermalisation is triggered has been left unanswered in previous studies. In
this work we obtain a precise estimate of $T_c$ as a function of $\kg$ and
substantiate our theoretical prediction through detailed numerical simulations. 

\subsection{Onset time}
Before we present our theoretical prediction for $T_c$, it is important at this 
stage to provide a more precise definition of this time.

\begin{figure}
\begin{center}
\includegraphics[width=1.0\columnwidth]{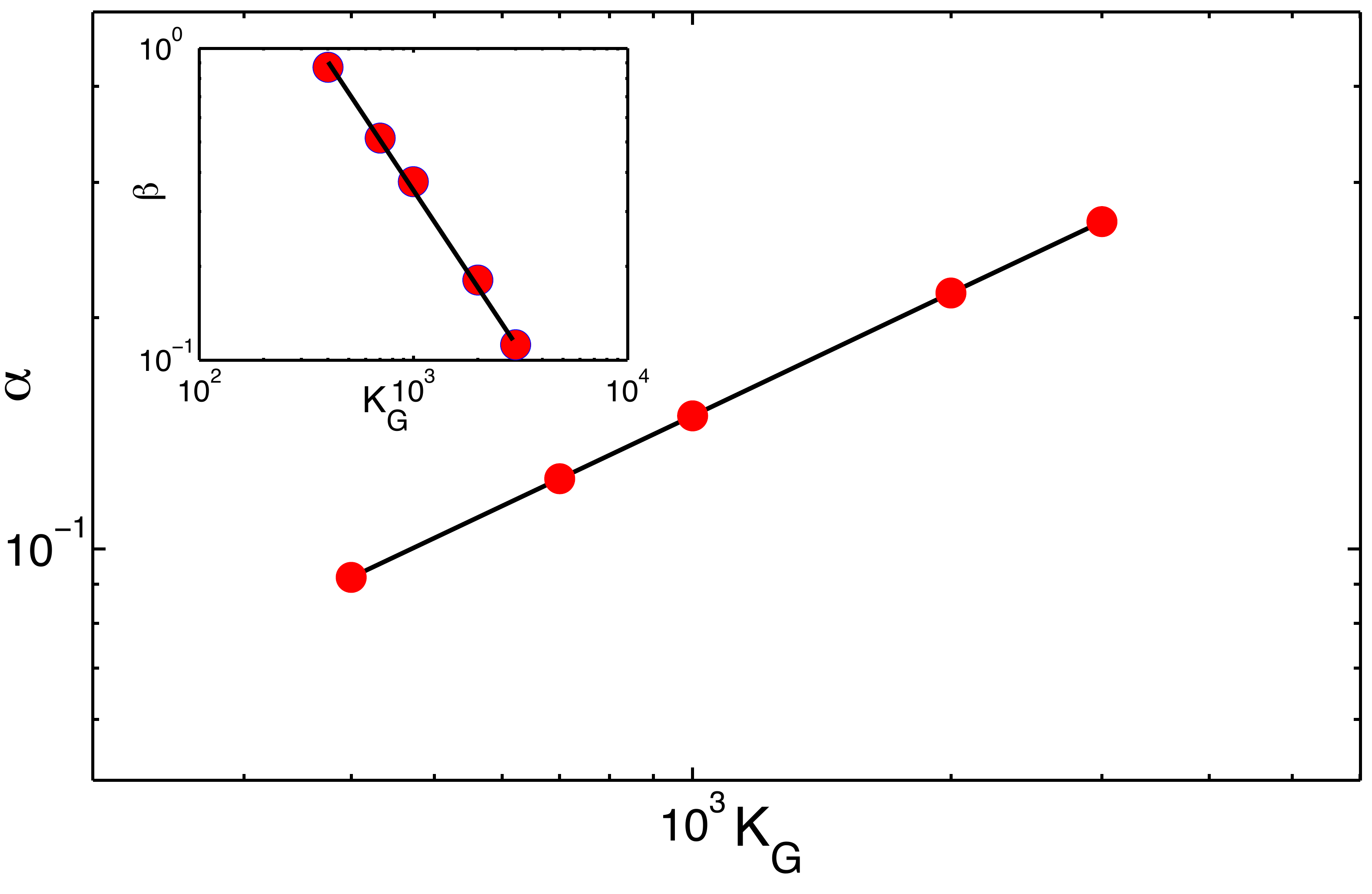}
\end{center}
\caption{(Log-log plots of the tyger amplitudes $\alpha$ and widths $\beta$ (inset), shown by large 
red dots, versus $\kg$ at a fixed 
time $t = 1.05$ from our simulations. The thick black line, both in the inset and the main figure, correspond to the theoretical 
prediction of $\beta \sim \kg^{-1}$ and $\alpha \sim \kg^{1/2}$, respectively. We find remarkable agreement between our 
analytical prediction and numerical data. We note that for reasons outlined in the text, 
we show data for values of $\kg$ for which $T_c > 1.05$.} 
\label{delta}
\end{figure}

Detailed numerical simulation show (Figs.~\ref{time-evolve} and \ref{bulge} as
well as in Ref.~\cite{ray11}) that in the early stages $t\gtrsim t_\star$ the
discrepancy $\Psi$  is small and localised at $x = x_{\rm tyger}$, where
$x_{\rm tyger}$ are points co-moving with the shock. With time this bulge
becomes bigger (still symmetric and localised), narrower (see
Fig.~\ref{bulge}), and the associated Reynolds stresses start increasing.  At a
critical time, the Reynolds stress becomes large enough to make the bulge
asymmetric.  This leads to a collapse of the bulge, accompanied by a spatial
spreading of the oscillations as well as the generation of different harmonics,
and the triggering of thermalisation of the system.  We define this critical
time $T_c$ as the time for onset of thermalisation. We note that numerically
$T_c$ is well-defined and easy to measure given that the asymmetry in the bulge
$\Psi$ can be determined clearly from the difference in the position of the
positive and negative peaks: This difference is zero for $t_\star \le t \le
T_c$ and becomes non-zero for $t > T_c$.

Before we proceed further, it is important at this stage to
define the nomenclature which we will use in obtaining our estimate for the
time of the onset of thermalisation.  As defined before, $t_\star =
\frac{1}{\rm {max} [\partial_x u]} \approx \frac{1}{\rm {max} [\partial_x v]}$
is the time when the complex singularity reaches the real domain. It was shown
that~\cite{ray11} tygers are born at a slightly earlier time $T_{\rm b}$ such
that $\tau_b \equiv t_\star - T_{\rm b} = \mathcal{O}\left (\kg^{-2/3}
\right)$. We define a new time scale $\tau_c \equiv  T_c - T_{\rm b}$ which
gives the estimate of the time scale for the onset of
thermalisation\footnote{As we shall see and explain later, for our numerical
simulations it is convenient to define $\tau_c = T_c - t_\star$.} and we obtain
theoretical results for the shifted time $\tau \equiv t - T_{\rm b}$.  It
should be noted that this is a natural choice for time since for $t < T_{\rm
b}$ tygers do not exist.  We will see below that $\tau_c$ indeed shows a
power-law behaviour in $\kg$ with a scaling form $\tau_c \sim \kg^{\xi}$; in
what follows we derive an explicit form for this new scaling exponent $\xi$ and
verify our theoretical predictions with data from detailed numerical
simulations.

\begin{table*}
\begin{center}
\begin{tabular}{|c|c|c|c|c|c|}
\hline
 & \multicolumn{2}{c}{ $\alpha \sim \tau^{\gamma_\alpha}\kg^{\delta_\alpha}$} &  \multicolumn{2}{c}{ $\beta \sim \tau^{\gamma_\beta}\kg^{\delta_\beta}$} & $\tau_c \sim \kg^{\xi}$ \\
\hline
\hline
 & $\gamma_\alpha$ & $\delta_\alpha$ & $\gamma_\beta$ & $\delta_\beta$ & $\xi$\\ 
\hline
Theory & 7/4 & 1/2 & -1 & -1 & $\xi = \frac{\delta_\beta - 2\delta_\alpha}{2\gamma_\alpha - \gamma_\beta}$ = -4/9\\
\hline
Simulations & 1.74 $\pm$ 0.04 & 0.50 $\pm$ 0.01 & -0.97 $\pm$ 0.08 & -1.01 $\pm$ 0.02 & -0.46 $\pm$ 0.07 \\
\hline
\end{tabular}
\end{center}
\caption{A summary of the new scaling exponents that we derive in this work. We see an excellent agreement, within error-bars, 
between our theoretical prediction and the exponents obtained, independently, from direct numerical simulations.}
\label{cases}
\end{table*}

We now turn our attention to the amplitude $\alpha$ and width $\beta$ of
the tyger.  Let the widths and amplitudes assume the scaling form $\beta \sim 
\tau^{\gamma_\beta}\kg^{\delta_\beta}$ and $\alpha \sim 
\tau^{\gamma_\alpha}\kg^{\delta_\alpha}$, respectively. It is important to note that the 
scaling ans\"atz introduce here for the widths and amplitudes of the bulge (tyger) at any time $\tau \equiv t - T_b$ is 
consistent with the result introduced in the previous section (and proved in~\cite{ray11}), namely the width and amplitude of the tyger 
at $t_\star$. This is because at $t = t_\star$, $\tau = \tau_b \sim \kg^{-2/3}$ (see~\cite{ray11}) and thence the scaling 
relation of the previous section follows from this ans\"atz. This is an important check of self-consistency of the 
theory. The asymmetry in the bulge 
occurs when the gradient of the Reynolds stresses  
become order one at time $\tau_c$. The Reynolds stress is defined as $\overline{\Psi^2}$,
where the overline indicates the typical, \textit{mesoscopic} average (spatial) ($\alpha$) over 
lengthscale larger than the Galerkin wavelength but smaller than other 
macroscopic scales in the problem; thence, dimensionally, the gradient of the Reynolds stress, namely 
$\frac{\alpha^2}{\beta}$, follows, since the relevant length scale over which 
this gradient should be taken $\sim \beta$. By using the assumed scaling form for $\alpha$ and $\beta$  
one obtains the scaling form for the time of the onset of thermalisation as
\begin{equation}
\tau_c \sim \kg^{\xi} \sim \kg^{\frac{\delta_\beta - 2\delta_\alpha}{2\gamma_\alpha - \gamma_\beta}}.
\label{prediction}
\end{equation}
We thus obtain the first theoretical estimate for the onset time of thermalisation in a truncated 
equation of idealised hydrodynamics and obtain a new scaling exponent for the same, namely, 
\begin{equation}
\xi = \frac{\delta_\beta - 2\delta_\alpha}{2\gamma_\alpha - \gamma_\beta}.
\label{xi}
\end{equation}
It now behooves us to determine, self-consistently, 
the exponents $\gamma_\alpha$, $\gamma_\beta$, $\delta_\alpha$, and $\delta_\beta$ and verify 
our predictions from detailed numerical simulations. 
 
\begin{figure}
\begin{center}
\includegraphics[width=1.0\columnwidth]{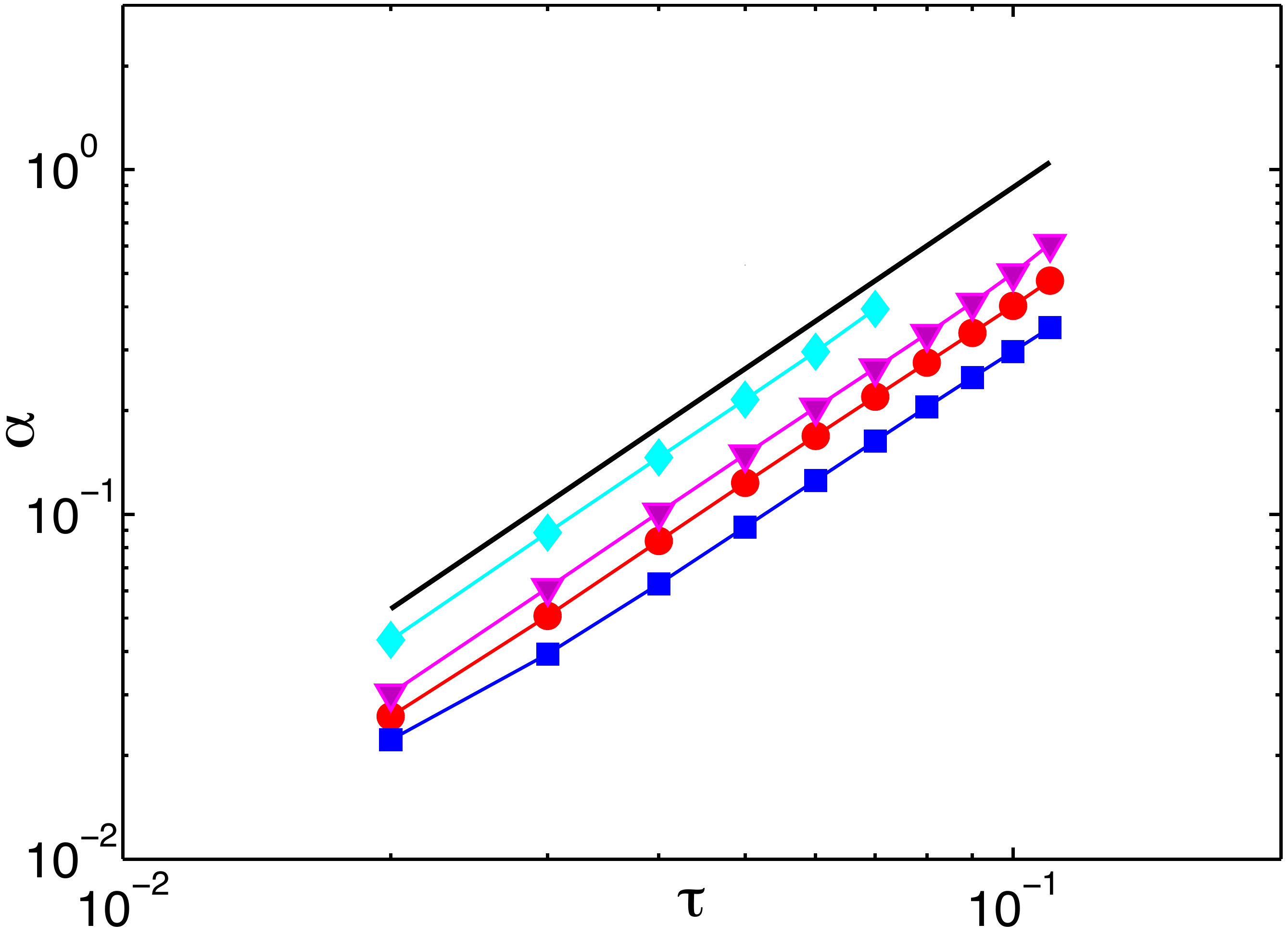}
\end{center}
\caption{Log-log plots of the tyger amplitudes $\alpha$, for various values of the truncation wavenumber $\kg$, 
as a function of the shifted time $\tau$. The different set of symbols correspond to different values of $\kg$, namely, 
blue filled squares for $\kg = 400$; big red dots for $\kg = 700$; filled pink triangles for $\kg = 1000$; and 
filled cyan diamond for $\kg = 2000$.  For each value of $\kg$, our theoretical prediction $\alpha \sim \tau^{7/4}$, shown via 
thick black lines, seems to be in excellent agreement with the numerical data.}
\label{gamma}
\end{figure}

As we know that the Galerkin-truncated Burgers equation conserves energy for
all time while displaying spatio-temporal chaotic behaviour at time $t \gg
t_\star$. This is unlike the case of the entropy or untruncated solution $u$ of
\eqref{burgers} which dissipate energy $\epsilon_{\rm shock}$ through the shock
for time $t \ge t_\star$. Therefore for the truncated equation to conserve
energy, $\epsilon_{\rm shock}$ gets transferred to the tygers via a resonant-wave-particle mechanism. 
Given that the estimate of energy contained in the tyger is $\alpha^2\beta$, this implies 
that for different values of $\kg$, the energy content of the tyger must be the same. This immediately suggests that 
for any finite $\tau$, the integral $\int_0^\tau \alpha^2\beta dt$ is independent of $\kg$. By using this argument 
we obtain the relation 
\begin{equation}
2\delta_\alpha + \delta_\beta = 0.
\label{delta-relation}
\end{equation}

Spatially the tygers are confined, due to resonance, to a region of width $w$.
For the tyger to grow, coherently, in a time interval $\tau$, phase mixing
constrains this region to be of an extent such that the velocity difference
across $w$ is of the order of $\lambda_G/\tau$.  Since $\lambda_G = 2\pi/\kg$
and the velocity difference across $w$ is proportional to $w$, this implies
that 
\begin{equation} 
\beta \sim \frac{1}{\tau\kg} 
\end{equation} 
yielding the exponents $\gamma_\beta = \delta_\beta = -1$. Furthermore, from
~\eqref{delta-relation}, we obtain $\delta_\alpha = 1/2$.

\begin{figure}
\begin{center}
\includegraphics[width=1.0\columnwidth]{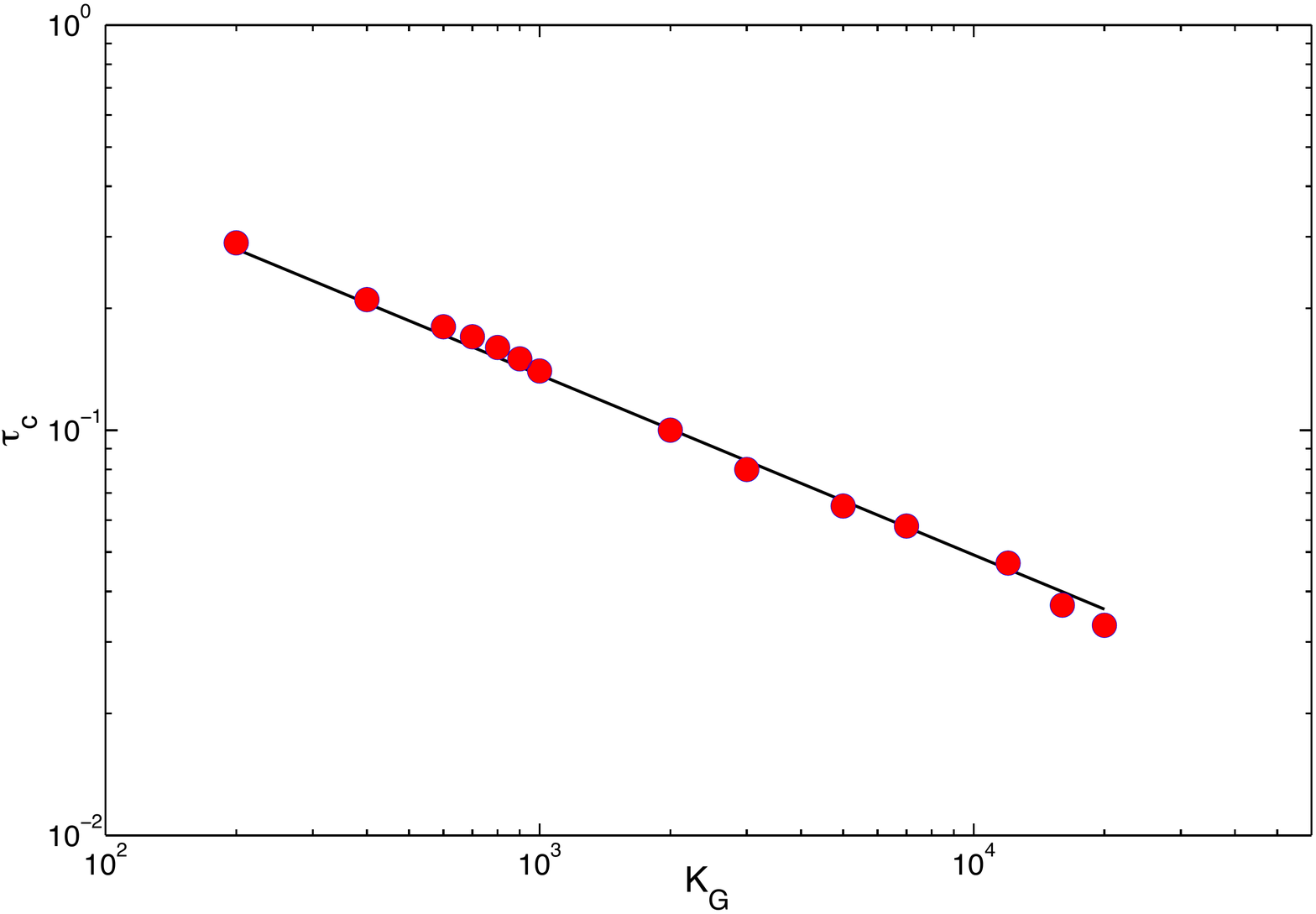}
\end{center}
\caption{Log-log plot of the onset time $\tau_c$ for thermalisation versus $\kg$. The data from our simulations is 
shown by big red dots; the thick black line connecting the data points is the theoretical prediction $\tau_c \sim \kg^{-4/9}$. 
Note the excellent agreement between our data and the theoretical prediction for smaller values of $\kg$ as compared to the 
larger values of $\kg$. Possible reasons for this discrepancy is discussed in the text.}
\label{tauc}
\end{figure}  

In Fig.~\ref{delta} we show plots of $\alpha$ and $\beta$ (inset) at a given time $t = 1.05$ as a function 
of $\kg$. The thick black line connecting our data from numerical simulations (shown as large red dots) 
correspond to the theoretical prediction of $\delta_\alpha \sim \kg^{1/2}$ and $\delta_\beta \sim 1/\kg$.
The figure shows not only a clear scaling, for both,  but also an excellent agreement of our numerical data with our 
theoretical predictions as well as validating our assumptions in arriving at our analytical results. We note that the 
range of $\kg$ shown in this figure does not extend all the way to our full range (e.g., see Fig.~\ref{tauc}). This is because 
at the time $t = 1.05$, for which the data is shown, solutions to the truncated equation with $\kg$ greater than 3000 have already 
thermalised (see Fig.~\ref{tauc}); hence the tygers have already collapsed for such wavenumbers at $t = 1.05$ and thus 
the measurement of their amplitudes and widths do not make sense. Our choice of $t = 1.05$ is motivated by having a time reasonably 
larger than $t_\star$ for which we still have a large range in $\kg$ to clearly illustrate the theoretically predicted scaling behaviour. 

Finally let us determine $\gamma_\alpha$. Let us consider $t = t_\star$, which, as we 
have noted before and proved in Ref.~\cite{ray11}, implies $\tau \sim \kg^{-2/3}$. 
At time $t_\star$, the untruncated equation shows a cubic root singularity. This implies, that 
because of Galerkin truncation the energy lost in the shock -- and hence gained as $\alpha^2\beta$ in 
the tyger -- is estimated as $\int_0^{\lambda_G} x^{2/3}dx \sim \kg^{-5/3}$. Setting $\tau = \kg^{-2/3}$, and 
using our previously obtained estimate $\gamma_\beta = \delta_\beta = -1$ and $\delta_\alpha = 1/2$, this 
suggests that 
\begin{equation}
\kg^{\frac{-4\gamma_\alpha + 2}{3}} \sim \kg^{-5/3}.
\end{equation} 
Hence we obtain $\gamma_\alpha = 7/4$.

From our simulations we calculate $\alpha$ for different values of $\kg$ and show plots $\alpha$ vs $\tau$ for representative 
values of $\kg$ in Fig.~\ref{gamma}. The thick black line, which is $\propto \tau^{7/4}$ 
is our theoretical prediction that for a given value of $\kg$, the amplitude of the tyger (upto the time of collapse) scales 
as $\alpha \sim \tau^{7/4}$. As in Fig.~\ref{delta}, we find excellent agreement between data obtained from our simulations 
with the theoretical prediction. 

Having obtained theoretically, and validated through simulations (Figs. ~\ref{delta} and \ref{gamma}), the values of the scaling 
exponents of the amplitude and width of the tygers before they collapse, let us once again return to the issue of the scaling 
of the onset time of thermalisation $\tau_c$. We had obtained before the scaling form of $\tau_c$ in Eq.~\ref{prediction}. We now 
use the values of the various exponents obtained thence to show, from Eq.~\ref{xi}, that $\xi = -4/9$, implying 
that thermalisation is triggered at a time
\begin{equation}
\tau_c \sim \kg^{-4/9}.
\label{prediction2}
\end{equation}
A summary of all these exponents is given in Table 1.

We now turn to our numerical simulations and obtain, for different values of
$\kg$, the time of collapse $\tau_c$. In our numerical
simulations we actually measure $\tau_c$ by using $t_\star$ (instead of $T_b$)
as the reference time, i.e., $\tau_c \equiv T_c - t_\star$. The reason for this
is because for the simulations we wanted a unique reference time $t_\star$
which is independent of $\kg$ (unlike $T_b$). This also reduces significantly
any measurement error in estimating $T_b$. Such a definition of $\tau_c$, for
our numerical simulations, is justified because for the values of $\kg$ used in
our simulations, $T_b$ and $t_\star$ are extremely close to each
other~\cite{ray11} and there is an order of magnitude separation between the
time scales $\tau_b$ and $\tau_c$.

In Fig.~\ref{tauc}, we show (red dots) the data obtained from our simulations.
The thick black line corresponds to the theoretical
prediction~\eqref{prediction2}. We find excellent agreement between our
analytical prediction and the data obtained from simulations.
Despite this confirmation of our theoretical predictions, it
is important to note that at extremely large values of $\kg$ we see a
noticeable discrepancy between the theoretical result and our data. The reasons
for this are two-fold: Firstly, as $\kg$ becomes larger and larger, $\tau_c$
become smaller and smaller. Hence a very accurate measurement of $\tau_c$,
numerically, becomes harder  because the relative error between the temporal
resolution of our simulations and $\tau_c$ become larger.  A second reason for
this is that numerically, for reasons mentioned before, we measure $\tau_c$
relative to $t_\star$ and not $T_b$. Hence for large $\kg$ when $\tau_c$
becomes smaller, our neglecting of $\tau_b$  (although there is an order of
magnitude scale separation between $\tau_c$ and $\tau_b$) probably starts
yielding a correction which is less neglible.

\section{Conclusions}

In this paper we provide the first prediction, analytically and validated by
numerically simulations, of the time when thermalisation is triggered in
finite-dimensional inviscid equations of hydrodynamics and hence solves a very
important problem in the interface of turbulence and statistical mechanics. We
show that, through the Galerkin-truncated Burgers model, thermalisation is
triggered on a time scale which decays as a power-law in $\kg$ with an exponent
which has been derived analytically by us and verified through numerical
simulations.   

Our results throw up important implications beyond the obvious realms of
non-equilibrium statistical physics. This has to do with using numerical
simulations for tracing complex singularities, in ideal equations of
hydrodynamics, by using the method of the analyticity strip~\cite{frisch83}. In
recent years (we refer the reader to Ref.~\cite{brachet} for the most recent
results and to Ref.~\cite{blow-up} for a recent review of results on finite-time
blow-ups via numerical simulations), with the advance of computing power, the
search for evidence for or against finite-time blow-up of the three-dimensional
Euler equations through numerical simulations have gained ground. As shown by
Bustamante and Brachet~\cite{brachet}, the temporal measurement of the
distance, to the real domain, of the nearest singularity, is limited not only
by computing power but also by the onset of thermalisation. Hence an estimate
of the time when thermalisation sets in will be have an important bearing on
interpreting the accuracy of measurements of complex singularities in time from
spectral, and hence Galerkin-truncated, simulations of the Euler equations. We
note in passing that the limitation in extrapolating in time the temporal
evolution of the width of the analyticity strip has been noted, amongst others,
in Ref.~\cite{brachet}.

There are of course several important questions which still remain unanswered.
Foremost amongst them is the need to see, numerically, if a similar scaling
argument holds for the incompressible three-dimensional Euler equation. This is
a massively challenging task even with the modern day computers. Secondly the
onset of thermalisation is necessarily accompanied by the generation of Fourier
harmonics other than $\kg$ which eventually lead to a white noise velocity
field with a flat spectrum. The precise mechanism of this is yet to be
understood in an analytical way. Furthermore, it has been observed
(Fig.~\ref{bulge} as well as in Ref.~\cite{ray11}) that just prior to and after
$\tau_c$, secondary bulges, reminiscent of the beating effect in acoustics,
develop on either side of the tyger. A systematic theory which explains the
full transition to thermalised states should capture this effect.  These
questions, and many more, are left for future work.

We thank Aritra Kundu for many useful discussions.
Both authors acknowledge financial support from the AIRBUS Group
Corporate Foundation Chair in Mathematics of Complex Systems established in
ICTS-TIFR. SSR also acknowledges the support from  DST (India) project
ECR/2015/000361 and the Indo-French Center for Applied Mathematics (IFCAM).
The direct numerical simulations were done on {\it Mowgli} at the ICTS-TIFR, Bangalore, India.

%%%%%%%%%%%%%%%%%%%%%%%%%%%%%%%%%%%%%%%%%%%%%%%%%%%%%%%%%%%%%%%%
%%%%%%%%%%%%%%%%%%%%%%%%%%%%%%%%%%%%%%%%%%

%%%%%%%%%%%%%%%%%%%%%%%%%%%%%%%%%%%%%%%%%%%%%%%%%%%%%%%%%%%%%%%%
\end{document}